\documentclass[twocolumn,secnumarabic,amssymb, nobibnotes, aps, prd]{revtex4-1}

\usepackage{graphicx}% Include figure files
\usepackage{dcolumn}% Align table columns on decimal point
\usepackage{bm}% bold math
\usepackage{slashed}
\usepackage{booktabs} % Top and bottom rules for table
\usepackage[font=small,labelfont=bf]{caption} % Required for specifying captions to tables and figures
\usepackage{amsfonts, amsmath, amsthm, amssymb} % For math fonts, symbols and environments
\usepackage{wrapfig} % Allows wrapping text around tables and figures

\begin{document}

%\preprint{AIP/123-QED}

\title{Low energy Eta-baryon interaction}% Force line breaks with \\

\author{M. G. L. Nogueira-Santos}
 \email{magwwo@gmail.com}%Lines break automatically or can be forced with \\
\author{C. C. Barros, Jr}%
 \email{barros.celso@ufsc.br}
\affiliation{ 
Departamento de F{\'{i}}sica, CFM, Universidade Federal de Santa Catarina\\ Florian{\'{o}}polis SC, CEP 88010-900, Brazil%\\This line break forced with \textbackslash\textbackslash
}%

\date{\today}% It is always \today, today,
             %  but any date may be explicitly specified

\begin{abstract}
  The $\eta$-Baryon interactions at low energies  are studied
in a model based in effective
chiral Lagragians that take into account baryons of spin 1/2 and spin 3/2 in the intermediate states. The interacting baryons to be considered in this work are
$B= N, \Lambda, \Sigma, \Xi$. We calculate the expected total and differential cross sections, phase-shifts and polarizations in the center-of-mass frame of reference for each reaction.
\end{abstract}

 % mesmo do pion-hyperon

\maketitle

%----------------------------------------------------------------------------------------
%	ARTICLE CONTENTS
%----------------------------------------------------------------------------------------

\section{Introduction}

The study of the meson-baryon interactions at low energies is a key element in order
to understand the strong interactions. Even if we only think in terms of the basic interactions
that determine the phenomenology of particle physics, this kind of interaction
may be considered of fundamental importance. Besides this point of view, these interactions
are the ones responsible for the structure and behavior of a countless number of physical
systems.

In previous works we have studied pion-hyperon and other kinds of meson-hyperon interactions
 \cite{BarrosJr2001}-\cite{sant4} and determined several elements of these processes (such as coupling constants and
cross-sections). In this work we will continue these studies, now considering the
$\eta$-baryon ($\eta B$) interactions.

When studying nuclear Physics, this kind of interaction is important, and is an element that
must be considered in order to obtain an accurate description of the proprieties of the
studied nuclei. The formation of $\eta$ bound states with nuclei \cite{Hayano, Bass}
and
the the $\eta$ production in interactions of nuclei with other particles \cite{kelkar, turin} are
interesting examples. An extension of these ideas is the inclusion of hypernuclei
inside this framework, and then, the knowledge of $\eta$-hyperon physics will be necessary. 

In the study of high energy collisions this kind of process also may be considered
in the description of the final-state interactions, when these systems reach
a state composed of interacting hadronic matter, in a way similar to the one shown
in \cite{Barros2011,Barros2008}. Despite the fact that the produced particles
reach high  momenta values in the laboratory frame, inside the medium the relative energy
is small, so, the
low energy interactions are the most important ones to be considered to explain
the observable effects.

These are just few examples of systems where the effects of $\eta B$ interactions should be
studied, but obviously, a much greater number of systems have important contributions of these
interactions.
So, in this work, we will study the low energy $\eta B$ interactions in a way that may be considered
a continuation of the works presented in \cite{BarrosJr2001}-\cite{sant4}, with a model based on
chiral effective
Lagrangians that consider baryons with spin 1/2 and 3/2 in the intermediate states of
the processes.

This work will show the following content: in Sec. II we will present the basic formalism
of the model and evaluate the analytical expressions for the considered amplitudes. In Sec.
III we will determine the coupling constants that appears in our expressions and show
the numerical results for some typical observables. In Sec. IV we will present the conclusions of this wor
k, and an Appendix will show some expressions that appear in the paper.

%------------------------------------------------

\section{Isosinglet Interation}

In this section we present the basic formalism
that will be used in order 
to calculate the Feynman diagrams for the
$\eta$-baryon interactions in processes with 
 baryons of spin 1/2 or spin 3/2 in the intermediate states.
 We will also consider a parametrization for the $\sigma$-meson exchange, as it has been done
 previously  \cite{BarrosJr2001}-\cite{sant2}.

In this work we will study four kinds of $\eta B$
interactions, $\eta N$, $\eta\Lambda$, $\eta\Sigma$ and $\eta\Xi$. For this purpose, we will take into
account effective chiral Lagrangians, already well studied in many works, as for example for the
$\pi N$ \cite{manc}-\cite{pi2}, $\pi Y$\cite{BarrosJr2001}-\cite{BarrosJr2006}, $KY$ \cite{sant, sant2, sant4} and
for the $\pi\Lambda_b$ \cite{sant3} interactions.
As the $\eta$-meson has isospin zero, all kinds of
$\eta B$ interactions  to be considered
to construct the scattering amplitudes
are isosinglet ones, so the use of the formalism for isospin states projection
operators will not be needed.

As the spin determines the form of the Lagrangian \cite{manc}-\cite{pi2} to be used in order to formulate the
model, 
we will consider two types of
chiral Lagrangians, that will represent spin-1/2 and spin-3/2  particles
in the intermediate state, given by
\begin{equation}
\label{eq1}
\mathcal{L}_{\eta BB_r}= \frac{g_{\eta BB_r}}{2m}\big(\overline{B}_r\gamma_\mu\gamma_5B\big)\partial^\mu\eta \ ,
\end{equation}

\begin{equation}
\label{eq2}
\mathcal{L}_{\eta BB^*}=g_{\eta BB^*}\overline{B}^{*\mu}\Big(g_{\mu\nu}+(Z+1/2)\gamma_\mu\gamma_\nu\Big)B\partial^\nu\eta \ ,
\end{equation}
where $B_r$, $B^*$ and $\eta$ are the baryon with spin 1/2, the baryon with
spin 3/2, and the $\eta$-meson fields with masses $m_r$, $m_{*}$ and $m_\eta$, respectively. $B$ is the incoming baryon with spin 1/2 and mass $m$. The parameter $Z$ represents the possibility of the
off-shell-$B^*$ having spin 1/2, and $g_{\eta BB_r}$ and
$g_{\eta BB^*}$ are the coupling constants. 

The Lagrangian formalism permits us to calculate the diagrams and then to find the scattering
amplitude $T_{\eta B}$, that may be decomposed in terms of the
 $A_{B'}$ and $B_{B'}$ amplitudes that for an interaction with intermediate state $B'$ is given by 
\begin{equation}
T_{\eta B}=\overline{u}(\vec{p'})\Big[A_{B'}+\frac{1}{2}(\slashed{k}+\slashed{k}')B_{B'}\Big]u(\vec{p}) \ ,
\label{eq3}
\end{equation}
where $\eta B$ represents the initial baryon with spin 1/2 ($B=N, \Lambda, \Sigma, \Xi$) and the  $\eta$-meson with spin 0. $u(\vec{p})$ and $\overline{u}(\vec{p'})$ are the spinors that represent the
initial and final baryons with momenta $\vec{p}$ and $\vec{p'}$. $\vec k$  and $\vec{k'}$ are the inital
and final momenta of the meson.

We also may decompose the scattering amplitude into   spin-non-flip and spin-flip amplitudes $f(k,x)$ and $g(k,x)$, that are useful in order to
to calculate the observables, then we write
\begin{equation}
T_{\eta B}=8\pi \sqrt{s}\big[f(k,x)+ i g(k.x)\ \vec{\sigma}.\hat{n}\big] \ ,
\end{equation}
where $\vec{\sigma}$ represents the Pauli matrices, $\hat{n}$ is an unitary vector in
the direction perpendicular to the scattering plane, $k=|\vec{k}|$, $x=\cos\theta$, $\theta$
is the scattering angle in the center-of-mass frame and $\sqrt{s}$ is the total energy as defined in the
Appendix.

These amplitudes may be expanded in terms of
partial-wave amplitudes,
\begin{equation}
f(k,x)=\sum_{l=0}^\infty{\Big[(l+1)a_{l+}^U(k)+la_{l-}^U(k)\Big] P_l (x)} \ ,
\label{eq5}
\end{equation}
\begin{equation}
g(k,x)=\sum_{l=1}^\infty{\Big[a_{l-}^U(k)-a_{l+}^U(k)\Big] P_l^{(1)} (x)}\ .
\label{eq6}
\end{equation}
where  $P_l(x)$ and $P_l^{(1)}(x)$ are Legendre's functions. The partial-wave amplitudes,
$a_{l\pm}^U(k)$,  are unitarized by the K-matrix method  \cite{BarrosJr2001}-\cite{BarrosJr2006},
\begin{equation}
a_{l\pm}^U=\frac{a_{l\pm}}{1-ika_{l\pm}} \ .
\end{equation}
This process is needed because the resulting  amplitudes are real when calculated at the tree level,
consequently they violate the unitarity of the $S$ matrix.   
The amplitudes $a_{l\pm}(k)$ may be determined by using the orthogonality relations for the Legendre
functions
\begin{eqnarray}
a_{l\pm}(k)=\frac{1}{2}\int_{-1}^1\Big[P_l(x)f_1(k) +P_{l\pm 1}(x)f_2(k)\Big] dx \ ,
\label{eq8}
\end{eqnarray}
where	$f_1(k)$ and $f_2(k)$ are functions defined in terms of the $A_{B'}$ and $B_{B'}$ amplitudes that
are calculated from the diagrams to be considered
\begin{equation}
f_1(k)=\frac{(E+m)}{8\pi \sqrt{s}}\big[A_{B'}+(\sqrt{s}-m)B_{B'}\big] \ ,
\end{equation}
\begin{equation}
f_2(k)=\frac{(E-m)}{8\pi \sqrt{s}}\big[-A_{B'}+(\sqrt{s}+m)B_{B'}\big] \ ,
\end{equation}
where $E$ is the baryon ($B$) energy in the center-of-mass frame. The processes that will be considered
in this work are shown in Fig. 1, and as we are interested in the low energy behavior of the reactions,
the $S$ and $P$ waves will dominate the scattering amplitudes and the amplitudes
relative to the larger values of the angular momentum will be just small corrections.

%%%%%%%%%%%%%%%%%%%%%%%%%%%%%%%%%%%%%%%%%%%%%%%%%%%%%%%%%%%%%%%
\begin{figure}[!htb]
 \includegraphics[width=0.45\textwidth]{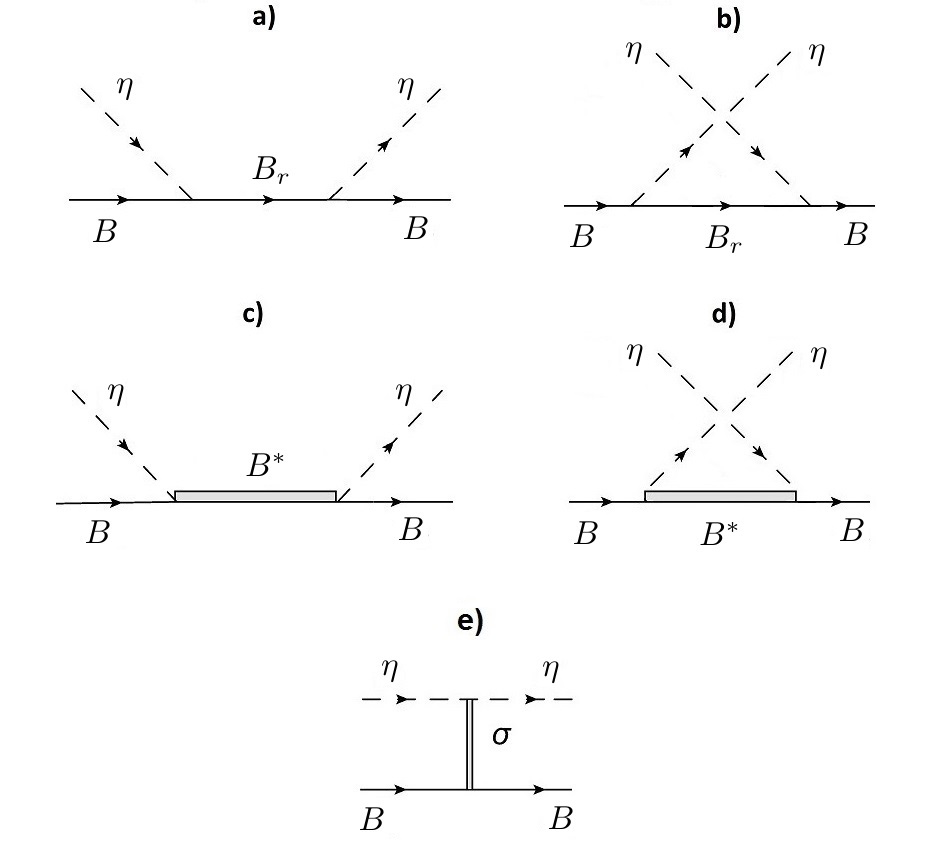}
\caption{Diagrams for the $\eta B$ interaction}\label{fig1}
\end{figure}
%%%%%%%%%%%%%%%%%%%%%%%%%%%%%%%%%%%%%%%%%%%%%%%%%%%%%%%%%%

As we can notice from the previous expressions,
the basic problem is to determine the $f(k, x)$ and $g(k, x)$ amplitudes, that will be used to obtain
the observables. For this purpose we calculate
the $A_{B'}$ and $B_{B'}$ amplitudes from the considered diagrams as the
resulting scattering
amplitude has the form of eq. (\ref{eq3}).
For the diagrams {\bf a)} and {\bf b)} shown in FIG.\ref{fig1}, 
that take into account a baryon spin 1/2 ($B'=B_r$) in the
intermediate state, considering the Lagragian (\ref{eq1}), we find
\begin{eqnarray}
\label{eq11}
A_{B_r}=\frac{g_{\eta BB_r}^2}{4m^2}(m_r+m)\bigg(\frac{s-m^2}{s-m_r^2}+\frac{u-m^2}{u-m_r^2}\bigg)\ ,
\end{eqnarray}
\begin{eqnarray}
\label{eq12}
B_{B_r}=\frac{g_{\eta BB_r}^2}{4m^2}\bigg[\frac{2m(m+m_r)+u-m^2}{u-m_r^2}\nonumber\\
-\frac{2m(m+m_r)+s-m^2}{s-m_r^2}\bigg]\ ,
\end{eqnarray}
where $s$, $t$ and $u$ are Mandelstam variables (the definition may be found in the Appendix).  

Making the same procedure for the diagrams {\bf c)} and {\bf d)} representing a spin 3/2
particle in the intermediate state ($B'=B^*$) and using the Lagrangian (\ref{eq2}), we have
\begin{eqnarray}
\label{eq13}
A_{B^*}=\frac{g_{ \eta BB^*}^2}{3m}\Bigg[\frac{\nu_*}{\nu_*^2-\nu^2}\hat{A}+m(a_0+a_z k.k')\Bigg]\ ,
\end{eqnarray}
\begin{eqnarray}
\label{eq14}
B_{B^*}=\frac{g_{\eta BB^*}^2}{3m}\Bigg[\frac{\nu}{\nu_*^2-\nu^2}\hat{B}+2m^2b_0\nu\Bigg]\ ,
\end{eqnarray}
 where $\nu_*$, $\nu$, $k.k'$ are defined in the Appendix and
\begin{eqnarray}
\hat{A}&=&\frac{(m_{*}+m)^2-m_\eta^2}{2m_{*}^2}\Big[2m_{*}^3-2m^3\nonumber\\
&&-2m m_{*}^2-2m^3-2m m_{*}^2\nonumber\\
&&-2m^2m_{*}+(2m-m_{*})m_\eta^2\Big]\nonumber\\
&&+\frac{3}{2}(m+m_{*})t\ ,
\end{eqnarray}
\begin{eqnarray}
\hat{B}&=&\frac{1}{2m_{*}^2}\Big[(m_{*}^2-m^2)^2-2m m_{*}(m+m_{*})^2\nonumber\\
&&-2m_\eta^2(m+m_{*})^2+6m_\eta^2m_{*}(m+m_{*})+m_\eta^4\Big]\nonumber\\
&&+ \frac{3}{2}t \ ,
\end{eqnarray}
\begin{eqnarray}
a_0=-\frac{(m +m_{*})}{m_{*}^2}\Big[2m_{*}^2+m m_{*}-m^2+2m_\eta^2\Big] \ ,
\label{eq:}
\end{eqnarray}
\begin{eqnarray}
a_z=\frac{4}{m_{*}^2}\Big[(m_{*}+m)Z+(2m_{*}+m)Z^2\Big]\ ,
\label{eq:}
\end{eqnarray}
\begin{eqnarray}
b_0=-\frac{4Z^2}{m_{*}^2}\ ,
\label{eq:}
\end{eqnarray}
 where $m_\eta$ is the $\eta$-meson mass. 

The last diagram  ${\bf e)}$, is an exchange of a scalar $\sigma$ meson calculated by the  parametrization \cite{BarrosJr2001}-\cite{BarrosJr2006}, \cite{leut1}-\cite{r1} 
\begin{eqnarray}
&&A_\sigma=a+bt\ ,\\
\label{eq32}
&&B_\sigma=0\ ,
\label{eq33}
\end{eqnarray}
where $a=1,05 m_\pi^{-1}$ and $b=-0,8m_\pi^{-3}$ are constants,
determined in terms of the pion mass $m_\pi=140 MeV$ \cite{pdg}.

\begin{table}[!htb]
\begin{ruledtabular}
  \begin{tabular}{ccccc} 
$$ &$ N$ & $\Lambda$ & $\Sigma$ &$\Xi$  \\ \hline
$a (m_\pi^{-1})$  & 1.05 & 0.54 & 0.47 & 0.19\\ 
$b (m_\pi^{-3})$  & -0.8 & 0.036 & 0.032 & 0.0074
 \end{tabular}
 \caption{Constants of scalar $\sigma$ meson parametrization}\label{tb1}
\end{ruledtabular}
\end{table}
%------------------------------------------------
%%%%%%%%%%%%%%%%%%%%%%%%%%%%%%%%%%%%%%%%%%%%%%%%%%%%%%%%%%%%%%%%%%%%%%%%%%%%%%%%%%%%%%%%%%%%%%%%%%%%%%%%%%%%%%%%%%%%%%%%%%%%%%%%%%%%%%%%%%%%%%%%%%%%%%%%%%%%%%%%%%%%%%%%%%%%%%%%%%%%%%%%%%%%%%
\section{Results of the $\eta B$ interactions}
%%%%%%%%%%%%%%%%%%%%%%%%%%%%%%%%%%%%%%%%%%%%%%%%%%%%%%%

In order to obtain numerical results for our calculations the coupling constants
that are considered
in the expressions must be determined. So,
in this section we will use the $SU(3)$ symmetry and the Breit-Wigner expression for
each intermediate particle that appear in the diagrams for
the $\eta B$ interaction to obtain their values.
Then we will present some results for the observables of interest,
the total cross sections, differential cross sections, phase-shifts and
polarizations.
\begin{table}[!htb]
\begin{ruledtabular}
  \begin{tabular}{ccccc} 
$Interaction$ &$Intermediate$ & $J^\pi$ & $I$ &$Mass$ ($MeV$) \\ \hline
$\eta N$ & $N$ & $1/2^+$&1/2&938\\ 
$\eta N$ & $N(1535)$& $1/2^-$&1/2&1535\\
$\eta N$ & $N(1650)$& $1/2^-$&1/2&1650\\
$\eta N$ & $N^*(1700)$& $3/2^-$&1/2 &1700\\  
$\eta N$ & $N(1710)$ &$1/2^+$&1/2 &1710 \\
$\eta \Lambda$ & $\Lambda$ & $1/2^+$&0&1116\\ 
$\eta \Lambda$ & $\Lambda(1670)$& $1/2^-$&0&1670\\
$\eta \Lambda$ & $\Lambda(1800)$& $1/2^-$&0&1800\\
$\eta \Sigma$ & $\Sigma$ & $1/2^+$&1&1116\\ 
$\eta \Sigma$ & $\Sigma(1750)$& $1/2^-$&1&1750\\
$\eta \Xi$ & $\Xi$ & $1/2^+$&1/2&1320
 \end{tabular}
 \caption{Intermediate baryons considered in the $\eta B$ interactions}\label{tb1}
\end{ruledtabular}
\end{table}
 
The $\eta B$ interactions to be studied in this work
are $\eta N$, $\eta \Lambda$, $\eta \Sigma$ and $\eta \Xi$.
The particles considered in the 
intermediate states for each $\eta B$ interaction are shown in Tab. \ref{tb1} \cite{pdg},
that presents their spin and parity $J^\pi$, isospin $I$ and mass. These reactions may
be understood in terms of the results obtained in the previous section. For intermediate
states with spin $1/2$ and spin $3/2$ we used the expressions
(\ref{eq11}), (\ref{eq12}) and (\ref{eq13}), (\ref{eq14}) respectively. 

For the lowest energy baryons in the $\eta B$ interaction we used the $SU(3)$ symmetry to
determine the coupling constants, that provides the general expression given by
\cite{stri, swart}
\begin{equation}
g_{\eta BB}=g_{\eta_8 BB}\cos\theta_{ps}-g_{\eta_0 BB}\sin\theta_{ps}
\label{eq:}
\end{equation}
where $\eta_0$ represents the singlet state and $\eta_8$ the octet state and $\theta_{ps}$,
the pseudoscalar meson mixing angle.  
Then we have
\begin{eqnarray}
g_{\eta_8 NN}&=&\frac{1}{\sqrt{3}}(4\alpha-1)f'\ ,\\
g_{\eta_8 \Lambda\Lambda}&=&\frac{2}{\sqrt{3}}(\alpha-1)f'\ ,\\
g_{\eta_8 \Sigma\Sigma}&=&-\frac{2}{\sqrt{3}}(\alpha-1)f'\ ,\\
g_{\eta_8 \Xi\Xi}&=&-\frac{1}{\sqrt{3}}(2\alpha+1)f'\ ,
\label{eq:}
\end{eqnarray}
and for the singlet states for all the considered $\eta B$ interactions the coupling
constant is given by
\begin{eqnarray}
g_{\eta_0 BB}&=&\sqrt{\frac{2}{3}}(4\alpha-1)f'\ ,
\label{eq:}
\end{eqnarray}
where we considered the values
$f'=g_{NN\pi}=13.4$ \cite{erc}, $\alpha=0.244$ and $\theta_{ps}=-23^\circ$. 
%%%%%%%%%%%%%%%%%%%%%%%%%%%%%%%%%%%%%%%%%%%%%%%%%%%%%%%%%%%%%%%%%%
\begin{figure}[!htb]
 \includegraphics[width=0.50\textwidth]{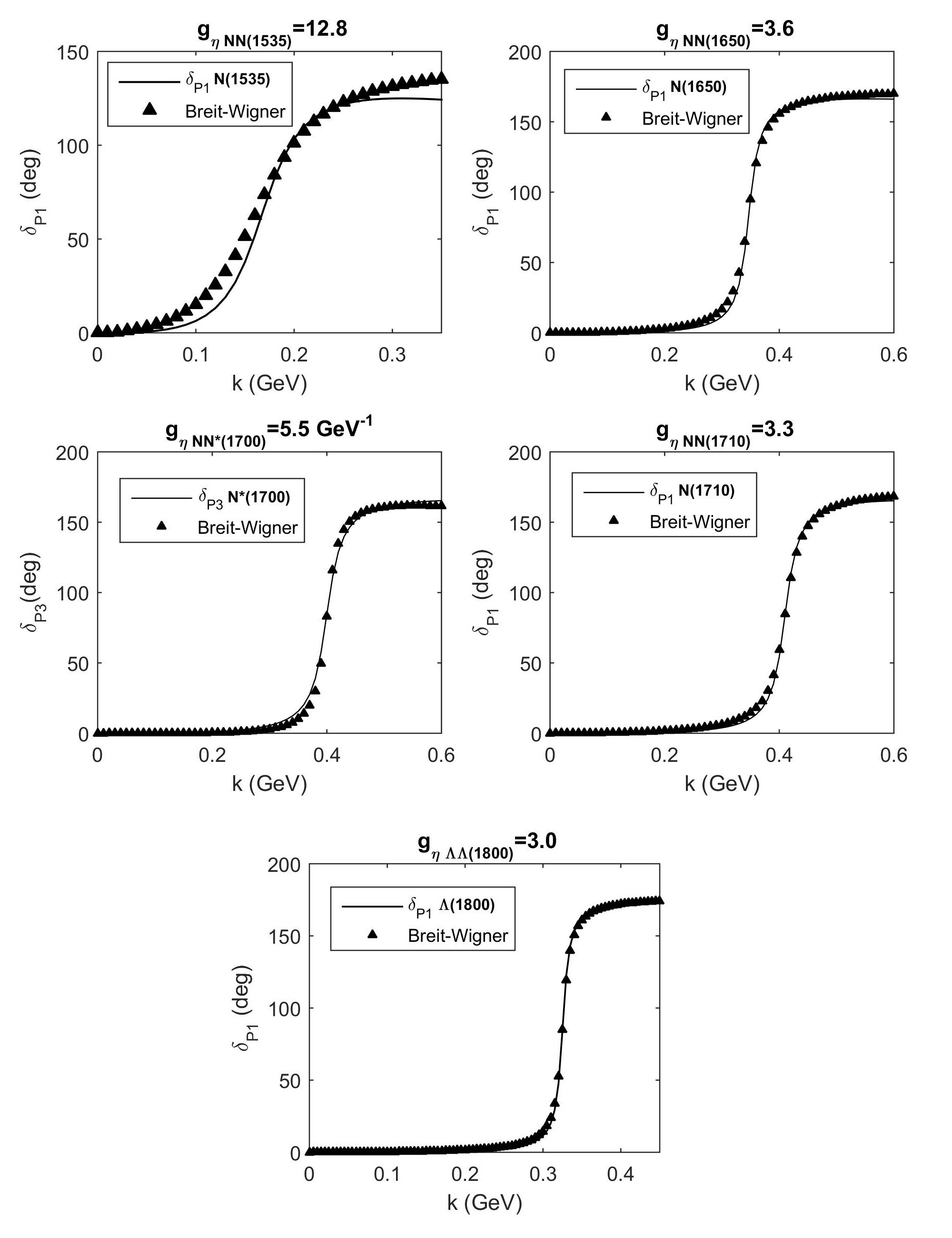}
\caption{Coupling constants adjustments for Breit-Wigner phase shifts for the $\eta B$ resonances }\label{fig2}
\end{figure}
%%%%%%%%%%%%%%%%%%%%%%%%%%%%%%%%%%%%%%%%%%%%%%%%%%%%%%%

To the calculate the coupling constants
for intermediates state with baryons of higher masses we use the relativistic Breit-Wigner expression
for each of them. In this method we compair the $\delta_{p1}$ or $\delta_{p3}$ phase shifts of the
resonant state with the Breit-Wigner expression
\begin{equation}
\delta_{l\pm}=\tan^{-1}\Bigg[\frac{\Gamma\Big(\frac{|\vec{k}|}{|\vec{k_0}|}\Big)^{2J+1}}{2(m'-\sqrt{s})}\Bigg] \ ,
\label{eq28}
\end{equation}
where $\Gamma$ is the Breit-Wigner width (experimental value that may be found in \cite{pdg})
and $|\vec{k}_0|$ is the center-of-mass momentum at the peak of the resonance and $m'$ its mass. 
%%%%%%%%%%%%%%%%%%%%%%%%%%%%%%%%%%%%%%%%%%%%%%%%%%%%%%%%%%%.
\begin{table}[!htb]
\begin{ruledtabular}
\begin{tabular}{ll}
$g_{\eta NN}$ & $-0.27$  \\
$g_{\eta NN(1535)}$ & $12.80$  \\
$g_{\eta NN(1650)}$ & $3.60$  \\
$g_{\eta NN^*(1700)}$ & $5.50$ $GeV^{-1}$ \\
$g_{\eta NN(1710)}$ & $3.30$  \\
$g_{\eta \Lambda\Lambda}$ & $-10.86$  \\
$g_{\eta \Lambda\Lambda(1670)}$ & $0.12$  \\
$g_{\eta \Lambda\Lambda(1800)}$ & $3.00$ \\
$g_{\eta\Sigma\Sigma}$ & $10.66$  \\
$g_{\eta \Sigma\Sigma(1750)}$ & $1.32$ \\
$g_{\eta\Xi\Xi}$ & $-10.70$  
\end{tabular}
\caption{Coumplig constants of the $\eta B$ interactions }\label{tb2}
\end{ruledtabular}
\end{table}
%%%%%%%%%%%%%%%%%%%%%%%%%%%%%%%%%%%%%%%%%%%%%%%%%%%%%%%%%%%%%%%%%%%%%%%
With this procedure we obtained the results shown in FIG.\ref{fig2}, where the
phase shifts calculated from the diagrams are compaired
with the ones determined by the
Breit-Wigner expression. The resulting coupling constants, for which
the best fits are achieved, are shown in the Tab. \ref{tb2}.

%%%%%%%%%%%%%%%%%%%%%%%%%%%%%%%%%%%%%%%%%%%%%%%%%%%%%%%
\begin{figure}[!htb]
 \includegraphics[width=0.5\textwidth]{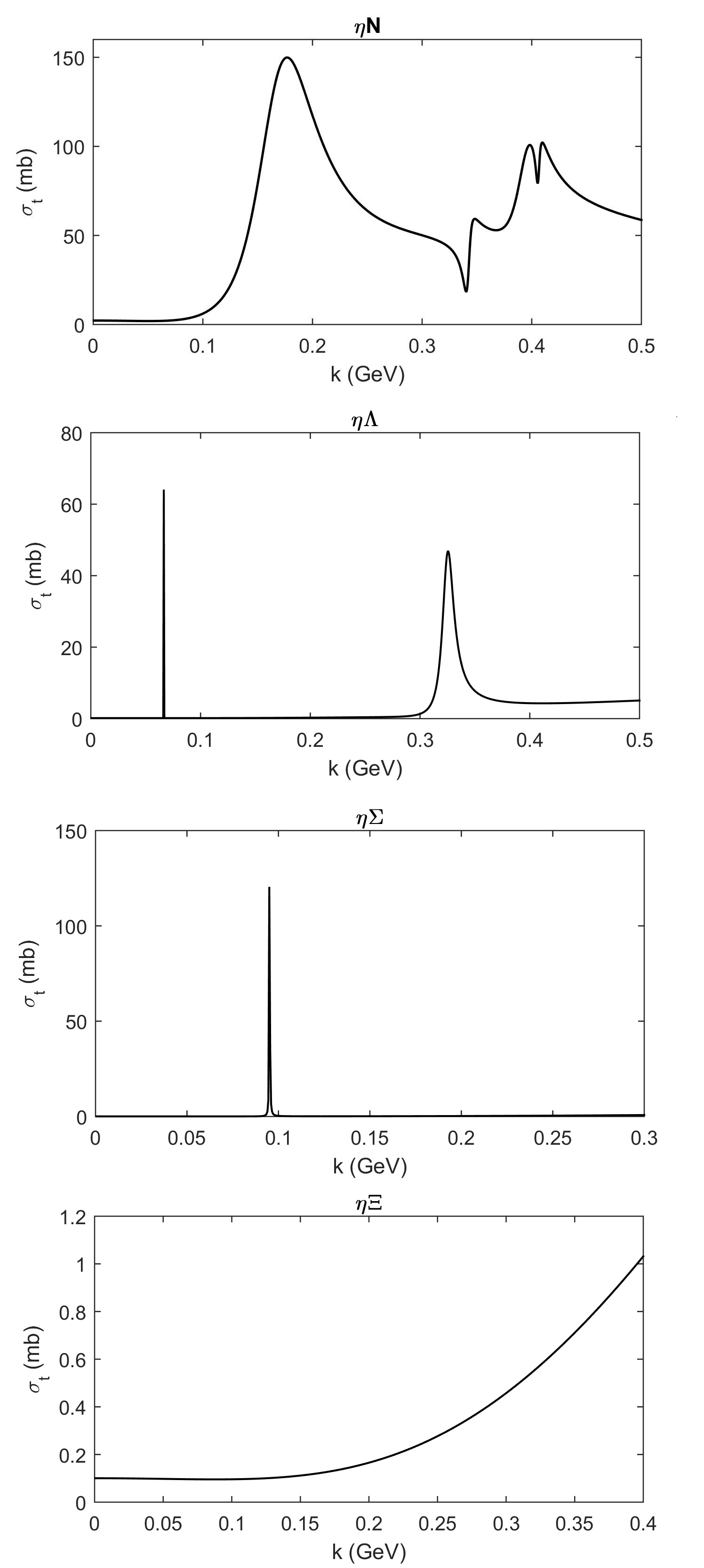}
\caption{Total cross sections for the $\eta B$ interactions}\label{fig3}
\end{figure}
%%%%%%%%%%%%%%%%%%%%%%%%%%%%%%%%%%%%%%%%%%%%%%%%%%%%%%%
\begin{figure}[!htb]
 \includegraphics[width=0.5\textwidth]{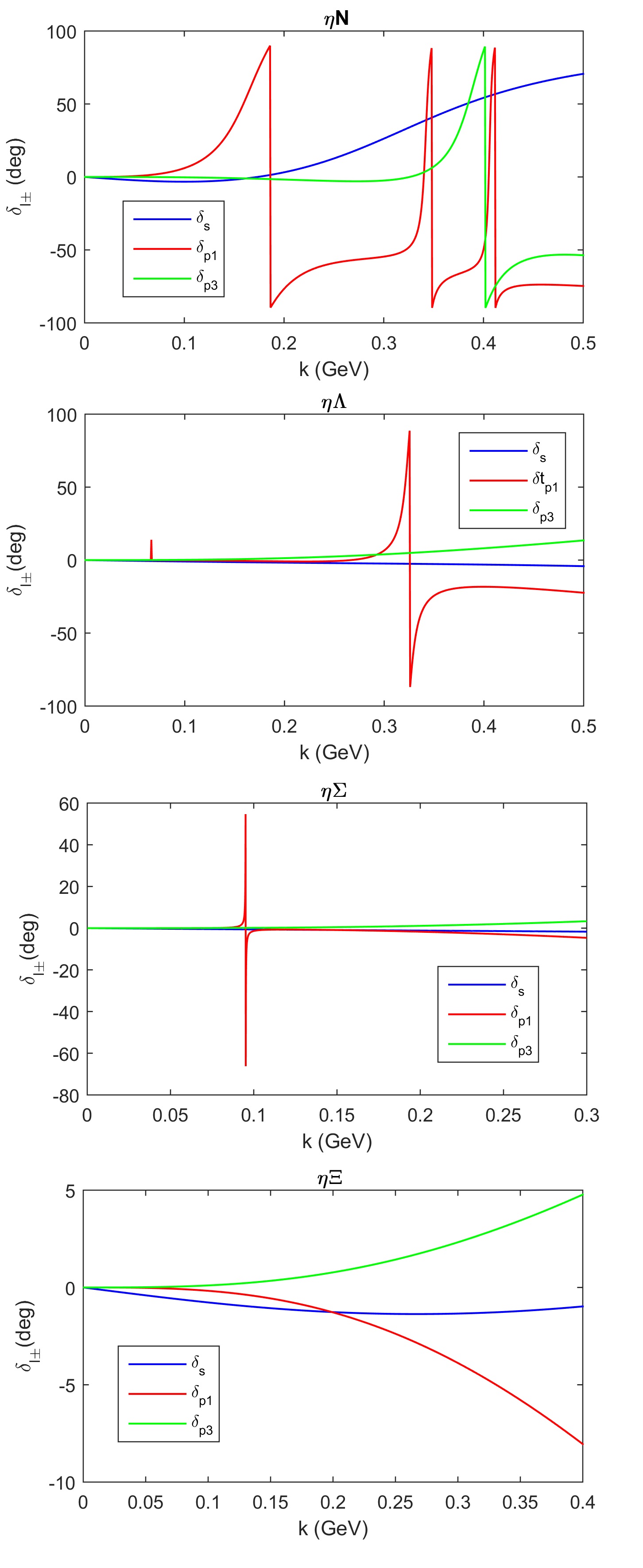}
\caption{Phase shifts for the $\eta B$ interactions}\label{fig4}
\end{figure}
%%%%%%%%%%%%%%%%%%%%%%%%%%%%%%%%%%%%%%%%%%%%%%%%%%%%%%%
\begin{figure}[!htb]
 \includegraphics[width=0.45\textwidth]{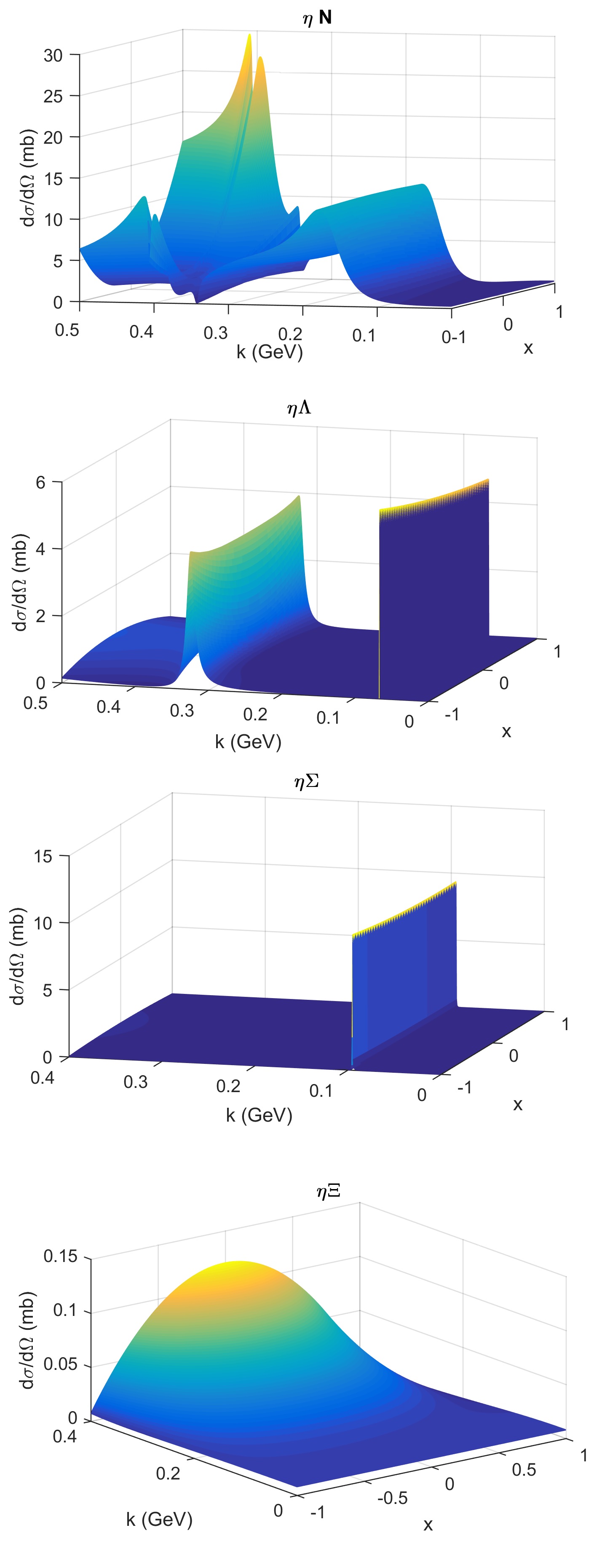}
\caption{Differential cross sections for the $\eta B$ interactions}\label{fig5}
\end{figure}
%%%%%%%%%%%%%%%%%%%%%%%%%%%%%%%%%%%%%%%%%%%%%%%%%%%%%%%
\begin{figure}[!htb]
 \includegraphics[width=0.45\textwidth]{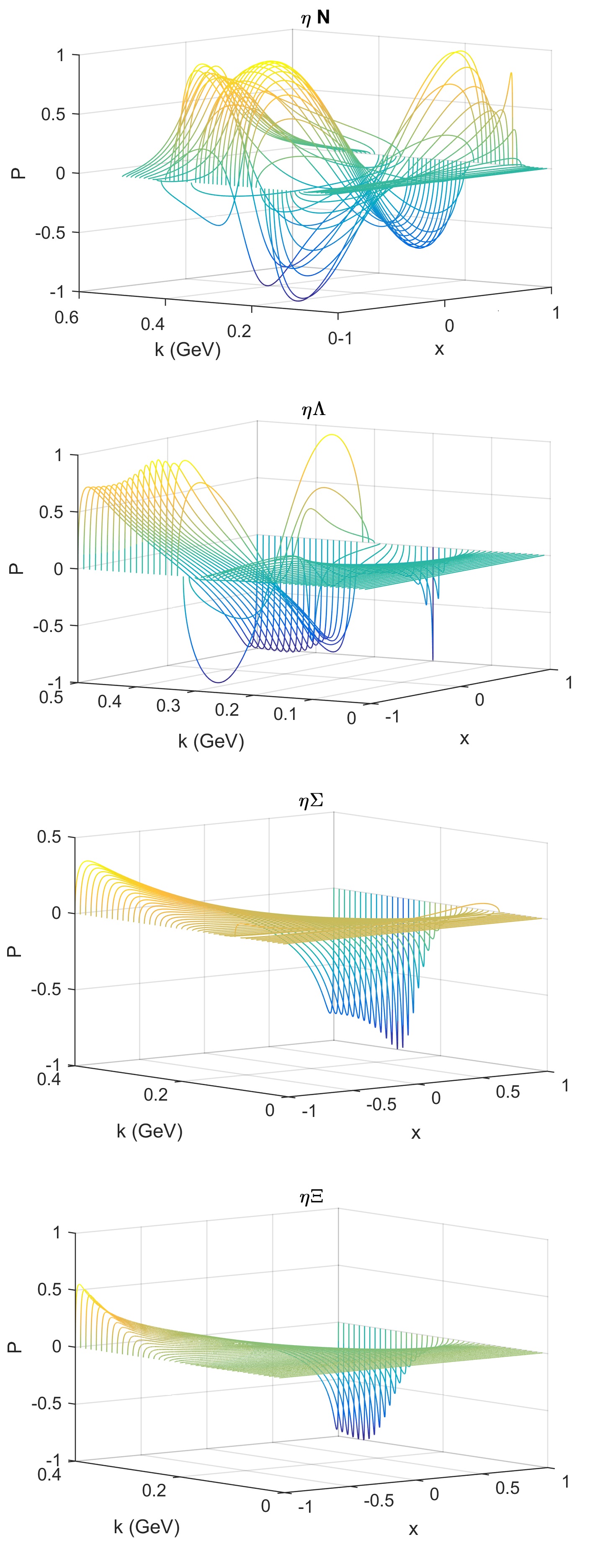}
\caption{Polarizations for the $\eta B$ interactions}
\label{fig6}
\end{figure}

The last step is to calculate the observables of each $\eta B$ interaction. By using the
unitarized amplitudes (\ref{eq5}) and (\ref{eq6}) in the center-of-mass frame, we construct the differential cross section
\begin{equation}
\frac{d\sigma}{d\Omega}=|f|^2+|g|^2\ ,
\label{eq:}
\end{equation}
 that integrated in the solid angle provides the total cross section
\begin{equation}
\sigma_T=4\pi \sum_l{\Big[(l+1)|a_{l+}^U|^2+l|a_{l-}^U|^2\Big]} \ .
\end{equation}
Also using the $f(k, x)$ and $g(k, x)$ amplitudes we have the polarization expression
\begin{equation}
\vec{P}=-2 \frac{Im(f*g)}{|f|^2+|g|^2}\hat{n} \ .
\label{eq:}
\end{equation}

The phase shifts are given in terms of the partial wave amplitudes, $a_{l\pm}$(\ref{eq8})
with the expression
\begin{equation}
\delta_{l\pm}=\tan^{-1}(ka_{l\pm}) \ .
\label{eq:}
\end{equation}

The results, considering $m_\eta=548 MeV$ \cite{pdg} and $Z=-0.5$, of the  observables of the $\eta N$, $\eta\Lambda$, $\eta\Sigma$ and $\eta\Xi$ interactions are show in FIG.\ref{fig3}, for the the
total cross sections, FIG.\ref{fig4}, for the phase shifts, FIG.\ref{fig5}, for the differential cross sections and in FIG.\ref{fig6} for the polarizations.

%%%%%%%%%%%%%%%%%%%%%%%%%%%%%%%%%%%%%%%%%%%%%%%%%%%%%%%%%%%%%%%%%%%%%%%%%%%%%%%%%%%%%%%%%%%%%%%%%%%%%%%%%%%%%%%%%%%%%%%%%
\section{Conclusions}

In this work we have studied the $\eta B$ interactions in a model based in chiral effective
lagrangians. The amplitudes of interest have been calculated, and then, we obtained all
the coupling constants that were needed. With these results it was possible
to calculate the phase shifts, total and differential cross sections and polarizations
for the low energy $\eta N$, $\eta \Lambda$, $\eta \Sigma$ and $\eta \Xi$  reactions.
As it has been pointed before, these results may be relevant when investigating
a large amount of physical systems.

As we can see in the figures, the cross sections follow the same pattern that may be found
in other meson-baryon interactions and are dominated by resonances at low energies. If we
imagine that in high energy collisions the observed baryons may be polarized by a final-state
interaction mechanism in a way similar to the one
that has been shown in \cite{Barros2011,Barros2008},
we may conclude that is possible to obtain baryons produced with significant polarization,
as far as, in the low energy processes, such as the ones shown in FIG. \ref{fig6}, under certain
conditions (of the momentum and of the scattering angle) the polarization may be large.

For the reasons exposed in this work, it is very important that this type of reaction
continues to be investigated and used in order to understand more complex systems.

%%%%%%%%%%%%%%%%%%%%%%%%%%%%%%%

\section{Acknowledgments}

This study has been partially supported by the Coordena\c c\~ao
de Aperfei\c coamento de Pessoal de N\'{\i}vel Superior
(CAPES) – Finance Code 001 and by CNPq.

%%%%%%%%%%%%%%%%%%%%%%%%%%%%%%%%%%%%%%%%%%%%%%%%%%%%%%%%%%%%%%%%%

%%%%%%%%%%%%%%%%%%%%%%%%%%%%%%%%%%%%%%%%%%%%%%%%%%%%%%%%%
\section{Appendix}
The Mandelstam variables are defined in terms of
 $p$ and $p'$, the initial and final baryon four-momenta, $k$ and $k'$ the initial and final meson four-momenta
\begin{equation}
s=(p+k)^2=(p'+k')^2=m^2+m_\eta^2+2Ek_0-2\vec{k}.\vec{p}\ ,
\label{eq:}
\end{equation}
\begin{equation}
u=(p'-k)^2=(p-k')^2=m^2+m_\eta^2-2Ek_0-2\vec{k}'.\vec{p}\ ,
\end{equation}
\begin{equation}
t=(p-p')^2=(k-k')^2=2(\vec{k})^2x-2(\vec{k})^2\ ,
\end{equation}
where, in the center-of-mass frame, the energies are
\begin{equation}
k_0=k'_0=\sqrt{(\vec{k})^2+m_\eta^2}\ ,
\end{equation}
\begin{equation}
E=E'=\sqrt{(\vec{k})^2+m^2}\ ,
\end{equation}
and the total momentum is null 
\begin{equation}
\vec{p}+\vec{k}=\vec{p}'+\vec{k}'=0\ .
\end{equation}
We also define, as usual 
\begin{equation}
x=\cos\theta\ ,
\end{equation}
where $\theta$ is the scattering angle in center-of-mass frame.

Other variables used in this work are
\begin{equation}
\nu_*=\frac{m_*^2-m^2-k.k'}{2m}\ ,
\end{equation}
\begin{equation}
\nu=\frac{s-u}{4m}=\frac{2Ek_0+(\vec{k})^2+(\vec{k})^2x}{2m}\ ,
\end{equation}
\begin{equation}
k.k'=m_\eta^2+(\vec{k})^2-(\vec{k})^2x=k_0^2-(\vec{k})^2x\ ,
\end{equation}
where $m$, $m_*$ and $m_\eta$ are the baryon mass, the resonance mass and the eta mass, respectively.

%----------------------------------------------------------------------------------------
%	REFERENCE LIST
%----------------------------------------------------------------------------------------
%%%%%%%%%%%%%%%%%%%%%%%%%%%%%%%%%%%%%%%%%%%%%%%%%%%%%%%%%%%%%%%%%%%%%%%%%%%%%%%%%%%%%%%%%%%%%%%%%%%%%%%%%%%%%%%%%%%%%%%%%%%%%%%%%%%%%%%%%%%%%%%%%%%%%%%%%%%%%%%%%%%%%%%%%%%%%

%----------------------------------------------------------------------------------------

\end{document}